\documentclass[journal ]{new-aiaa}
\usepackage[utf8]{inputenc}
\usepackage{textcomp}

\usepackage{graphicx}
\usepackage{subfigure}
\usepackage{amsmath}
\usepackage[version=4]{mhchem}
\usepackage{siunitx}
\usepackage{longtable,tabularx}
\title{Far-field approximations to the derivatives and integrals of the Green's function for the Ffowcs Williams and Hawkings equation}

\author{Zhiteng Zhou,Shizhao Wang\footnote{Corresponding author, wangsz@lnm.imech.ac.cn.}}
\affil{LNM, Institute of Mechanics, Chinese Academy of Sciences, Beijing,China, 100190}
\affil{School of Engineering Sciences, University of Chinese Academy of Sciences, Beijing,China, 100049}

\begin{document}

\maketitle

\begin{abstract}

We report far-field approximations to the derivatives and integrals of the Green's function for the Ffowcs  Williams and Hawkings equation in the frequency domain. The approximations are based on the far-field asymptotic of the Green's function. 
The details of the derivations of the proposed formulations are provided.
\end{abstract}

\section{Introduction}
The Ffowcs Williams and Hawkings (FW-H) equation \cite{lockard2000efficient} is an inhomogeneous wave equation that extents the Lighthill's acoustic analogy to the flow including moving/permeable boundaries. Sources in the FW-H equation consist of monopole, dipole and quadrupole terms \cite{ffowcs1969sound}. 
The solution to the FW-H equation can be expressed as surface integrals of monopole and dipole sources and a volume integral of the quadrupole sources. The quadrupole term is usually ignored under the assumption that the monopole and dipole terms dominate the far-field sound in very low Mach numbers flows. However, recent research has suggested that ignoring the quadrupole sources may result in spurious sound even at relatively low Mach numbers \cite{zhong2017sound, mao2020analysis}.

Wang et al. \cite{wang1996computation} proposed a surface correction to eliminate the spurious sound associated with the quadrupole source. The surface integral correction is then improved by \cite{nitzkorski2014dynamic} to account for the non-uniform convection velocity. Ikeda et al. \cite{ikeda2012quadrupole} and Lockard and Casper \cite{lockard2005permeable} proposed the surface corrections for the frequency-domain method. In particular, the surface correction proposed by Lockard and Casper \cite{lockard2005permeable} consists of a series of surface integrals. The spurious sound generated by flows is successfully estimated using the surface integral series. 
However, The surface correction involves computation of the high-order derivatives of the Green's function, which is quite complicated and nontrivial to be calculated \cite{ikeda2012quadrupole}. The 2nd order derivative was reported in \cite{gloerfelt2003direct} while the analytical formulations of higher-order ones are usually referred to the use of symbolic algebra packages \cite{lockard2005permeable}.

In the present work, we propose analytical formulations to approximate the derivatives of the Green's function at the far-field. Inspired by the derivation of the high-order derivatives, we also propose a simplified formulation for computing the integrals of the Green's function, which may be applied to evaluating local contribution to the quadrupole term. The mathematical induction method is then used to prove the proposed formulation. The remained parts of the letter is organized as follows. We will give the simplified formulations for the derivatives and multiple integrals of the Green's functions in Section II, prove the simplified high-order derivatives of the Green's function for 2D and 3D flows in Section III, prove the simplified integrals using mathematical induction method in Section IV and draw conclusions in Section V.

\section{Far-field approximations to the derivatives and integrals of the Green's function}
The derivatives and integrals of the Green's function for the frequency-domain FW-H equation can be approximated at the far field as follows
\begin{equation}\tag{1}
\begin{split}   
\frac{{{\partial }^{l}}}{\partial y_{1}^{l}}(\frac{{{\partial }^{2}}G{_{2D}}^{{}}(\mathbf{x};\mathbf{y})}{\partial {{y}_{i}}\partial {{y}_{j}}})\approx \left( \frac{\partial \varphi_{2D} (\mathbf{x};\mathbf{y})}{\partial {{y}_{1}}} \right)^{l}\frac{{{\partial }^{2}}G{_{2D}}^{{}}(\mathbf{x};\mathbf{y})}{\partial {{y}_{i}}\partial {{y}_{j}}},
\end{split}
\end{equation}
\begin{equation}\tag{2}
\begin{split}   
\frac{{{\partial }^{l}}}{\partial y_{1}^{l}}(\frac{{{\partial }^{2}}G_{^{3D}}^{{}}(\mathbf{x};\mathbf{y})}{\partial {{y}_{i}}\partial {{y}_{j}}})\approx \left( \frac{\partial \varphi_{3D} (\mathbf{x};\mathbf{y})}{\partial {{y}_{1}}} \right)^{l}\frac{{{\partial }^{2}}G_{^{3D}}^{{}}(\mathbf{x};\mathbf{y})}{\partial {{y}_{i}}\partial {{y}_{j}}},
\end{split}
\end{equation}
\begin{equation}\tag{3}
\begin{split}   
\frac{{{\partial }^{2}}G{^l_{2D}}(\mathbf{x};\mathbf{y})}{\partial {{y}_{i}}\partial {{y}_{j}}}\approx \left( \frac{\partial \varphi_{2D} (\mathbf{x};\mathbf{y})}{\partial {{y}_{1}}} \right)^{\text{-}l}\frac{{{\partial }^{2}}G{_{2D}}^{{}}(\mathbf{x};\mathbf{y})}{\partial {{y}_{i}}\partial {{y}_{j}}},
\end{split}
\end{equation}
\begin{equation}\tag{4}
\begin{split}   
\frac{{{\partial }^{2}}G_{^{3D}}^{l}(\mathbf{x};\mathbf{y})}{\partial {{y}_{i}}\partial {{y}_{j}}}\approx \left( \frac{\partial \varphi_{3D} (\mathbf{x};\mathbf{y})}{\partial {{y}_{1}}} \right)^{\text{-}l}\frac{{{\partial }^{2}}G_{^{3D}}^{{}}(\mathbf{x};\mathbf{y})}{\partial {{y}_{i}}\partial {{y}_{j}}}
\end{split}
\end{equation}
where
\begin{equation}\tag{5}
\begin{split}   
\varphi_{2D} (\mathbf{x};\mathbf{y})=i\left[ Mk({{x}_{1}}-{{y}_{1}})/{{\beta }^{2}}+\frac{\pi }{4}-\frac{k}{{{\beta }^{2}}}R \right],
\end{split}
\end{equation}
\begin{equation}\tag{6}
\begin{split}   
{{\varphi }_{3D}}(\mathbf{x};\mathbf{y})=-ik\frac{\left( d-M\left( {{x}_{1}}-{{y}_{1}} \right) \right)}{{\beta }^{2}},
\end{split}
\end{equation}
\begin{equation}\tag{7}
\begin{split}   
\frac{{{\partial }^{2}}{{G}^{n}}(\mathbf{x};\mathbf{y})}{\partial {{y}_{i}}\partial {{y}_{j}}}=\underbrace{\int_{\infty }^{{{y}_{1}}}{\int_{\infty }^{{{\xi }_{n}}}{\int_{\infty }^{{{\xi }_{n-1}}}{\cdots \int_{\infty }^{{{\xi }_{2}}}{{}}}}}}_{n}\frac{{{\partial }^{2}}G(\mathbf{x};{{\xi }_{1}},{{y}_{2}})}{\partial {{y}_{i}}\partial {{y}_{j}}}d{{\xi }_{1}}d{{\xi }_{2}}\cdots d{{\xi }_{n-1}}d{{\xi }_{n}}
\end{split}
\end{equation}

$G{_{2D}}$ is the asymptotic Green's function for 2D flows and $G{_{3D}}$ is the Green's function for 3D flows. $R=\sqrt{{{({{x}_{1}}-{{y}_{1}})}^{2}}+{{\beta }^{2}}{{({{x}_{2}}-{{y}_{2}})}^{2}}}$,
$\beta =\sqrt{1-{{M}^{2}}}$, $\text{ }k=\frac{\omega}{{c}_{o}}$ and
$d=\sqrt{{{({{x}_{1}}-{{y}_{1}})}^{2}}+{{\beta }^{2}}{{({{x}_{2}}-{{y}_{2}})}^{2}}+{{\beta }^{2}}{{({{x}_{3}}-{{y}_{3}})}^{2}}}$. $M$ is the Mach number of the freestream flow. The observer and source locations are denoted by $\mathbf{x}$ and $\mathbf{y}$, respectively.

\section{Derivatives of the Green's function at the far field}
We report the derivation of derivatives of the Green's function in this section.

To prove Eq. (1), we start from the far-field condition $|\mathbf{x}|>>|\mathbf{y}|$. By using $|\mathbf{x}|>>|\mathbf{y}|$, we have
\begin{equation}\tag{8}
\begin{split}
   & \frac{{{\partial }^{{{k}_{1}}}}R}{\partial y_{1}^{{{k}_{1}}}}\approx O(\frac{1}{{{R}^{{{k}_{1}}-1}}})\frac{{{\partial }^{{}}}R}{\partial y_{1}^{{}}} ,\\ 
   & \frac{{{\partial }^{{{k}_{1}}}}({{R}^{\text{-1/2}}})}{\partial y_{1}^{{{k}_{1}}}}\approx O(\frac{1}{{{R}^{{{k}_{1}}-1}}})\frac{{{\partial }^{{}}}{{R}^{\text{-1/2}}}}{\partial y_{1}^{{}}}, \\ 
\end{split}
\end{equation}
where $k{}_{1} \ge 1$. Taking the $k$th-order derivative of $G_{2D}$ with respect to ${{y}_{1}}$, we have
\begin{equation}\tag{9}
\begin{split}
   \frac{{{\partial }^{l}}{{G}_{2D}}(\mathbf{x};\mathbf{y})}{\partial y_{1}^{l}}=\frac{i}{4\beta }{{(\frac{2{{\beta }^{2}}}{\pi l})}^{1/2}}\sum\limits_{{{k}_{1}}=0}^{l}{C_{l}^{{{k}_{1}}}\frac{{{\partial }^{{{k}_{1}}}}{{\exp }^{\varphi (\mathbf{x};\mathbf{y})}}}{\partial y_{1}^{{{k}_{1}}}}}\frac{{{\partial }^{l-{{k}_{1}}}}({{R}^{-1/2}})}{\partial y_{1}^{l-{{k}_{1}}}},
\end{split}
\end{equation}
where $C_{l}^{{{k}_{1}}}$ is the binomial coefficient. By using the first line of Eq. (8), the ${{k}_{1}}$th-order derivative of ${{\exp }^{\varphi (\mathbf{x};\mathbf{y})}}$ can be estimated with ignoring the high-order derivatives of $\varphi (\mathbf{x};\mathbf{y})$ as follows
\begin{equation}\tag{10}
\begin{split}
   \frac{{{\partial }^{{{k}_{1}}}}{{\exp }^{\varphi (\mathbf{x};\mathbf{y})}}}{\partial y_{1}^{{{k}_{1}}}}\approx {{(\frac{{{\partial }^{{}}}\varphi (\mathbf{x};\mathbf{y})}{\partial y_{1}^{{}}})}^{{{k}_{1}}}}{{\exp }^{\varphi (\mathbf{x};\mathbf{y})}}\text{   }({{k}_{1}} \ge 1).
\end{split}
\end{equation}
Combination of Eqs. (9) and (10) results in Eq. (1) where we ignore the terms in the right-hand-side of Eq. (9) with $k_1<l$ according to the second line of Eq. (8). 

The derivation of Eq. (2) is similar to Eq. (1).

\section{Integrals of the Green's function at the far field}
We report the derivation of integrals of the Green's function in this section.

To prove Eq. (3), we start from proving the equations as follows with the mathematical induction method,

\begin{equation}\tag{11}
\begin{split}
   \frac{{{\partial }^{2}}G_{^{2D}}^{l}(\mathbf{x};\mathbf{y})}{\partial {{y}_{i}}\partial {{y}_{j}}}\approx \frac{{{\partial }^{2}}\left[ {{(\frac{\partial \varphi (\mathbf{x};\mathbf{y})}{\partial {{y}_{1}}})}^{-l}}G_{^{2D}}^{{}}(\mathbf{x};\mathbf{y}) \right]}{\partial {{y}_{i}}\partial {{y}_{j}}}.
\end{split}
\end{equation}
For $l=1$, Eq. (11) reduces to
\begin{equation}\tag{12}
\begin{split}
\frac{{{\partial }^{2}}G_{^{2D}}^{1}(\mathbf{x};\mathbf{y})}{\partial {{y}_{i}}\partial {{y}_{j}}}=\int_{\infty }^{{{y}_{1}}}{\left( {{\left. \frac{{{\partial }^{2}}G_{^{2D}}^{{}}(\mathbf{x};\mathbf{y})}{\partial {{y}_{i}}\partial {{y}_{j}}} \right|}_{{{y}_{1}}={{\xi }_{1}}}} \right)d{{\xi }_{1}}}.
\end{split}
\end{equation}
After transforming the partial derivative with respect to $\mathbf{y}$ to $\mathbf{x}$ and using integration by parts, the right-hand-side of Eq. (12) becomes
\begin{equation}\tag{13}
\begin{split}
  \frac{{{\partial }^{2}}G_{^{2D}}^{1}(\mathbf{x};\mathbf{y})}{\partial {{y}_{i}}\partial {{y}_{j}}}=&\frac{{{\partial }^{2}}}{\partial {{x}_{i}}\partial {{x}_{j}}}\int_{\infty }^{{{y}_{1}}}{\frac{i}{4\beta }{{(\frac{2{{\beta }^{2}}}{\pi kR})}^{1/2}}{{\exp }^{\varphi (\mathbf{x};{{\xi }_{1}},{{y}_{2}})}}d{{\xi }_{1}}} \\ 
  =&\frac{{{\partial }^{2}}}{\partial  {{x}_{i}}\partial {{x}_{j}}}\int_{\infty }^{{{y}_{1}}}{\frac{\partial }{\partial {{\xi }_{1}}}\left[ \frac{i}{4\beta }{{(\frac{2{{\beta }^{2}}}{\pi kR})}^{1/2}}{{(\frac{\partial \varphi (\mathbf{x};{{\xi }_{1}},{{y}_{2}})}{\partial {{\xi }_{1}}})}^{-1}}{{\exp }^{\varphi (\mathbf{x};{{\xi }_{1}},{{y}_{2}})}} \right]d{{\xi }_{1}}}- \\ 
  \rm{   }&\frac{{{\partial }^{2}}}{\partial {{x}_{i}}\partial {{x}_{j}}}\int_{\infty }^{{{y}_{1}}}{\left[ {{(\frac{\partial \varphi (\mathbf{x};{{\xi }_{1}},{{y}_{2}})}{\partial {{\xi }_{1}}})}^{-1}}(\mathbf{x};\mathbf{y}){{\exp }^{\varphi (\mathbf{x};{{\xi }_{1}},{{y}_{2}})}} \right]\frac{\partial }{\partial {{\xi }_{1}}}(\frac{i}{4\beta }{{(\frac{2{{\beta }^{2}}}{\pi kR})}^{1/2}})d{{\xi }_{1}}}. \\ 
\end{split}
\end{equation}
From Eq. (8), we know that the last term on the right-hand-side of Eq. (13) is ignorable compared with the left-hand-side of Eq. (13), thus we have 
\begin{equation}\tag{14}
\begin{split}
  \frac{{{\partial }^{2}}G_{^{2D}}^{1}(\mathbf{x};\mathbf{y})}{\partial {{y}_{i}}\partial {{y}_{j}}}\approx \frac{{{\partial }^{2}}}{\partial {{x}_{i}}\partial {{x}_{j}}}\int_{\infty }^{{{y}_{1}}}{\frac{\partial }{\partial {{\xi }_{1}}}\left[ \frac{i}{4\beta }{{(\frac{2{{\beta }^{2}}}{\pi kR})}^{1/2}}{{(\frac{\partial \varphi (\mathbf{x};{{\xi }_{1}},{{y}_{2}})}{\partial {{\xi }_{1}}})}^{-1}}{{\exp }^{\varphi (\mathbf{x};{{\xi }_{1}},{{y}_{2}})}} \right]d{{\xi }_{1}}}. 
\end{split}
\end{equation}
Considering the limitation of 2D Green’s function to infinity and transforming the partial derivative with respect to $\mathbf{x}$ back to $\mathbf{y}$, Eq. (14) becomes
\begin{equation}\tag{15}
\begin{split}
   \frac{{{\partial }^{2}}G_{^{2D}}^{1}(\mathbf{x};\mathbf{y})}{\partial {{y}_{i}}\partial {{y}_{j}}}&\approx \frac{{{\partial }^{2}}}{\partial {{y}_{i}}\partial {{y}_{j}}}\left[ \frac{i}{4\beta }{{(\frac{2{{\beta }^{2}}}{\pi kR})}^{1/2}}{{(\frac{\partial \varphi (\mathbf{x};\mathbf{y})}{\partial {{y}_{1}}})}^{-1}}{{\exp }^{\varphi (\mathbf{x};\mathbf{y})}} \right] \\ 
   & \text{                   =}\frac{{{\partial }^{2}}\left( {{(\frac{\partial \varphi (\mathbf{x};\mathbf{y})}{\partial {{y}_{1}}})}^{-1}}G_{^{2D}}^{{}}(\mathbf{x};\mathbf{y}) \right)}{\partial {{y}_{i}}\partial {{y}_{j}}}, \\ 
\end{split}
\end{equation}
which proves that Eq. (11) is valid when $l=1$. 

By using the mathematical induction method, we assume that Eq. (11) is valid when $l=h$
\begin{equation}\tag{16}
\begin{split}
\frac{{{\partial }^{2}}G_{^{2D}}^{h}(\mathbf{x};\mathbf{y})}{\partial {{y}_{i}}\partial {{y}_{j}}}\approx \frac{{{\partial }^{2}}\left[ {{(\frac{\partial \varphi (\mathbf{x};\mathbf{y})}{\partial {{y}_{1}}})}^{-h}}G_{^{2D}}^{{}}(\mathbf{x};\mathbf{y}) \right]}{\partial {{y}_{i}}\partial {{y}_{j}}}.
\end{split}
\end{equation}
According to Eqs. (7) and (16), we have
\begin{equation}\tag{17}
\begin{split}
\frac{{{\partial }^{2}}G_{^{2D}}^{h+1}(\mathbf{x};\mathbf{y})}{\partial {{y}_{i}}\partial {{y}_{j}}}\approx \int_{\infty }^{{{y}_{h}}}{\left( {{\left. \frac{{{\partial }^{2}}\left[ {{(\frac{\partial \varphi (\mathbf{x};\mathbf{y})}{\partial {{y}_{1}}})}^{-l}}G_{^{2D}}^{{}}(\mathbf{x};\mathbf{y}) \right]}{\partial {{y}_{i}}\partial {{y}_{j}}} \right|}_{{{y}_{1}}={{\xi }_{1}}}} \right)d{{\xi }_{1}}}. 
\end{split}
\end{equation}
Using integration by parts, we reform Eq. (17) as follows
\begin{equation}\tag{18}
\begin{split}
  \frac{{{\partial }^{2}}G_{^{2D}}^{h+1}(\mathbf{x};\mathbf{y})}{\partial {{y}_{i}}\partial {{y}_{j}}}\approx &\frac{{{\partial }^{2}}}{\partial {{x}_{i}}\partial {{x}_{j}}}\int_{\infty }^{{{y}_{1}}}{\left( {{\left. \frac{i}{4\beta }{{(\frac{\partial \varphi (\mathbf{x};\mathbf{y})}{\partial {{y}_{1}}})}^{-h}}{{(\frac{2{{\beta }^{2}}}{\pi kR})}^{1/2}}{{\exp }^{\varphi (\mathbf{x};\mathbf{y})}} \right|}_{{{y}_{1}}={{\xi }_{1}}}} \right)d{{\xi }_{1}}} \\ 
  =&\frac{{{\partial }^{2}}}{\partial {{x}_{i}}\partial {{x}_{j}}}\int_{\infty }^{{{y}_{1}}}{\left\{ {{\left. \frac{\partial }{\partial {{y}_{1}}}\left[ \frac{i}{4\beta }{{(\frac{\partial \varphi (\mathbf{x};\mathbf{y})}{\partial {{y}_{1}}})}^{-h}}{{(\frac{2{{\beta }^{2}}}{\pi kR})}^{1/2}}{{(\frac{\partial \varphi (\mathbf{x};\mathbf{y})}{\partial {{y}_{1}}})}^{-1}}{{\exp }^{\varphi (\mathbf{x};\mathbf{y})}} \right] \right|}_{{{y}_{1}}={{\xi }_{1}}}} \right\}d{{\xi }_{1}}}- \\ 
  &\frac{{{\partial }^{2}}}{\partial {{x}_{i}}\partial {{x}_{j}}}\int_{\infty }^{{{y}_{1}}}{\left\{ {{\left. \left[ {{(\frac{\partial \varphi (\mathbf{x};\mathbf{y})}{\partial {{y}_{1}}})}^{-1}}{{\exp }^{\varphi (\mathbf{x};\mathbf{y})}} \right]\frac{\partial }{\partial {{y}_{1}}}\left[ \frac{i}{4\beta }{{(\frac{\partial \varphi (\mathbf{x};\mathbf{y})}{\partial {{y}_{1}}})}^{-h}}{{(\frac{2{{\beta }^{2}}}{\pi kR})}^{1/2}} \right] \right|}_{{{y}_{1}}={{\xi }_{1}}}} \right\}d{{\xi }_{1}}}. \\ 
\end{split}
\end{equation}
From Eq. (8), we know that the last term on the right-hand-side of Eq. (18) is ignorable compared with the left-hand-side of the Eq. (18). Similar to the derivation of Eq. (15), Eq. (18) could be approximated by 
\begin{equation}\tag{19}
\begin{split}
  \frac{{{\partial }^{2}}G_{^{2D}}^{h+1}(\mathbf{x};\mathbf{y})}{\partial {{y}_{i}}\partial {{y}_{j}}}\approx \frac{{{\partial }^{2}}}{\partial {{y}_{i}}\partial {{y}_{j}}}\left[ \frac{i}{4\beta }{{(\frac{2{{\beta }^{2}}}{\pi kR})}^{1/2}}{{(\frac{\partial \varphi (\mathbf{x};\mathbf{y})}{\partial {{y}_{1}}})}^{-(h+1)}}{{\exp }^{\varphi (\mathbf{x};\mathbf{y})}} \right].
\end{split}
\end{equation}
Eq. (19) shows that Eq. (11) is valid when $l=h+1$. Thus, Eq. (11) is proved according to the mathematical induction method.

By using Leibniz product rule, the right-hand-side of Eq. (11) equals to
\begin{equation}\tag{20}
\begin{split}
 & \frac{{{\partial }^{2}}}{\partial {{y}_{i}}\partial {{y}_{j}}}\left[ \frac{i}{4\beta }{{(\frac{2{{\beta }^{2}}}{\pi k})}^{1/2}}\left( {{R}^{-1/2}}{{(\frac{\partial \varphi (\mathbf{x};\mathbf{y})}{\partial {{y}_{1}}})}^{-l}} \right){{\exp }^{\varphi (\mathbf{x};\mathbf{y})}} \right]= \\ 
 & \frac{i}{4\beta }{{(\frac{2{{\beta }^{2}}}{\pi k})}^{1/2}}\left[ {{R}^{-1/2}}{{(\frac{\partial \varphi (\mathbf{x};\mathbf{y})}{\partial {{y}_{1}}})}^{-l}}{{\exp }^{\varphi (\mathbf{x};\mathbf{y})}}\left( (\frac{\partial \varphi (\mathbf{x};\mathbf{y})}{\partial {{y}_{i}}})(\frac{\partial \varphi (\mathbf{x};\mathbf{y})}{\partial {{y}_{j}}})+o(1) \right) \right]\text{+} \\ 
 & \frac{i}{4\beta }{{(\frac{2{{\beta }^{2}}}{\pi k})}^{1/2}}\left[ \frac{\partial \left( {{R}^{-1/2}}{{(\frac{\partial \varphi (\mathbf{x};\mathbf{y})}{\partial {{y}_{1}}})}^{-l}} \right)}{\partial {{y}_{i}}}\frac{\partial \left( {{\exp }^{\varphi (\mathbf{x};\mathbf{y})}} \right)}{\partial {{y}_{j}}} \right]+ \\ 
 & \frac{i}{4\beta }{{(\frac{2{{\beta }^{2}}}{\pi k})}^{1/2}}\left[ \frac{\partial \left( {{R}^{-1/2}}{{(\frac{\partial \varphi (\mathbf{x};\mathbf{y})}{\partial {{y}_{1}}})}^{-l}} \right)}{\partial {{y}_{j}}}\frac{\partial \left( {{\exp }^{\varphi (\mathbf{x};\mathbf{y})}} \right)}{\partial {{y}_{i}}} \right]+ \\ 
 & \frac{i}{4\beta }{{(\frac{2{{\beta }^{2}}}{\pi k})}^{1/2}}\left[ \frac{{{\partial }^{2}}\left( {{R}^{-1/2}}{{(\frac{\partial \varphi (\mathbf{x};\mathbf{y})}{\partial {{y}_{1}}})}^{-l}} \right)}{\partial {{y}_{i}}\partial {{y}_{j}}}{{\exp }^{\varphi (\mathbf{x};\mathbf{y})}} \right]. \\ 
\end{split}
\end{equation}
In the far field, $\frac{\partial \varphi (\mathbf{x};\mathbf{y})}{\partial {{y}_{i}}}$ is of order $O(1)$ and the derivative of ${{R}^{-1/2}}{{(\frac{\partial \varphi (\mathbf{x};\mathbf{y})}{\partial {{y}_{1}}})}^{-l}}$ can be expressed by $o({{R}^{-1/2}}{{(\frac{\partial \varphi (\mathbf{x};\mathbf{y})}{\partial {{y}_{1}}})}^{-l}})$. Thus, the last three terms of Eq. (20) are ignorable compared with the first term on the right-hand-side. We ignore the derivatives corresponding to ${{(\frac{\partial \varphi (\mathbf{x};\mathbf{y})}{\partial {{y}_{1}}})}^{-l}}$ and obtain
\begin{equation}\tag{21}
\begin{split}
  \frac{{{\partial }^{2}}}{\partial {{y}_{i}}\partial {{y}_{j}}}\left[ \frac{i}{4\beta }{{(\frac{2{{\beta }^{2}}}{\pi k})}^{1/2}}\left( {{R}^{-1/2}}{{(\frac{\partial \varphi (\mathbf{x};\mathbf{y})}{\partial {{y}_{1}}})}^{-l}} \right){{\exp }^{\varphi (\mathbf{x};\mathbf{y})}} \right]\approx& 
  {{(\frac{\partial \varphi (\mathbf{x};\mathbf{y})}{\partial {{y}_{1}}})}^{-l}}\frac{{{\partial  }^{2}}}{\partial {{y}_{i}}\partial {{y}_{j}}}\left[ \frac{i}{4\beta }{{(\frac{2{{\beta }^{2}}}{\pi k})}^{1/2}}{{R}^{-1/2}}{{\exp }^{\varphi (\mathbf{x};\mathbf{y})}} \right]\\
  \text{=}&{{(\frac{\partial \varphi (\mathbf{x};\mathbf{y})}{\partial {{y}_{1}}})}^{-l}}\frac{{{\partial }^{2}}{{G}_{\text{2}D}}}{\partial {{y}_{i}}\partial {{y}_{j}}}.  
\end{split}
\end{equation}
Finally, substituting Eq. (21) in Eq. (19) yields Eq. (3). 

The derivation of Eq. (4) is similar to that of Eq. (3).

\section{Conclusion}

 The computation of the high-order derivatives of the Green's function for the FW-H equation are required in eliminating the spurious sound associated with the quadrupole sources. The computation of the high-order derivatives of the Green's function is usually quite complicated and nontrivial. In this work, we simplify the computation of the high-order derivatives at the far-field. In addition, we propose a simplified formulation for computing the integral of the Green's function. We use the far-field asymptotic Green’s function to compute its derivatives and integrals. The details of the derivations are reported.

\bibliography{sample}

\end{document}